\documentclass[aps,prl,twocolumn,groupedaddress,amsmath,amssymb, preprintnumbers]{revtex4-1}
\pdfoutput=1
\usepackage{graphicx}  % needed for figures
\usepackage{color}
\usepackage{hyperref} % hyperlinked references in PDF

\newcommand{\gsim}{\mathrel{\lower.8ex\vbox{\lineskip=.1ex\baselineskip=0ex \hbox{$>$}\hbox{$\sim$}}}}
\newcommand{\lsim}{\mathrel{\lower.8ex\vbox{\lineskip=.1ex\baselineskip=0ex \hbox{$<$}\hbox{$\sim$}}}}

\begin{document}

\preprint{UTTG-15-16}

\title{Secretly Asymmetric Dark Matter}
\author{Prateek Agrawal}
\affiliation{
   Department of Physics,\\
   Harvard University,\\
   Cambridge, MA 02138, USA
   }
\author{Can Kilic}
\affiliation{
   Theory Group, Department of Physics and Texas Cosmology Center\\
   University of Texas at Austin,\\
   Austin, TX 78712, USA
   }
\author{Sivaramakrishnan Swaminathan}
\affiliation{
   Theory Group, Department of Physics and Texas Cosmology Center\\
   University of Texas at Austin,\\
   Austin, TX 78712, USA
   }
\author{Cynthia Trendafilova}
\affiliation{
   Theory Group, Department of Physics and Texas Cosmology Center\\
   University of Texas at Austin,\\
   Austin, TX 78712, USA
   }
\date{\today}
\begin{abstract}
We study a mechanism where the dark matter number density today
arises from asymmetries generated in the dark sector in the early
universe, even though total dark matter number remains zero throughout the history of the universe. The dark matter
population today can be completely symmetric, with annihilation rates
above those expected from thermal WIMPs. We give a simple example
of this mechanism using a benchmark model of flavored dark matter. We
discuss the experimental signatures of this setup, which arise mainly
from the sector that annihilates the symmetric component of dark
matter.

 \end{abstract}

\maketitle

\noindent
Asymmetric dark matter (ADM)~\cite{Nussinov:1985xr, Gelmini:1986zz,
Barr:1990ca, Barr:1991qn, Kaplan:1991ah,  Kaplan:2009ag,
Petraki:2013wwa,Zurek:2013wia} is motivated by the observation that
the dark matter and baryon energy densities today are comparable, so
that for dark matter masses of a few GeV, the number densities of the
dark and visible sectors are also roughly comparable. The baryon
number density today is set by an asymmetry, which suggests that dark
matter could also be asymmetric, with the origin of the two
asymmetries being related.  In order to realize the conventional ADM
scenario, a mechanism has to be put in place in order to break
$U(1)_{\chi}$, a
symmetry which guarantees conservation of dark matter (DM) number, in
much the same way that baryon number has to be broken in order to
generate an asymmetry in the visible sector.

In this paper we study the possibility that for a dark
sector with multiple states, the ADM paradigm can be
realized without having to break
$U(1)_{\chi}$.
Asymmetries can be generated in the different dark sector states, while
keeping the total charge under the $U(1)_{\chi}$ at zero.
If heavier states in the dark sector
decay to lighter ones after DM annihilations have frozen out~\cite{Falkowski:2011xh,Hardy:2014dea}, then the
final DM population is in fact symmetric, even though its
abundance was set by an asymmetry. For this reason we will refer to
this mechanism as Secretly Asymmetric Dark Matter (SADM). The idea of repopulating the symmetric component of DM at late times through oscillations has also been explored~\cite{Buckley:2011ye,Cirelli:2011ac,Tulin:2012re,Okada:2012rm,Chen:2015yuz}.

The relic abundance of DM in this mechanism is in some ways similar to
the abundance of charged stable particles in the Standard Model (SM).
Even though the abundances of baryons and leptons are set by an
initial asymmetry, the universe is always charge neutral and
$U(1)_{\mathrm{EM}}$ is never broken. If protons were to decay at late
times, the universe could end up with a symmetric population of
electrons and positrons which is secretly asymmetric.

\vspace{0.1in} \noindent {\it The Generation of the Asymmetry:}
Flavored dark matter (FDM) models~\cite{MarchRussell:2009aq, Batell:2011tc,
Agrawal:2011ze, Kile:2013ola,  Kumar:2013hfa, Lopez-Honorez:2013wla, Batell:2013zwa, Agrawal:2014una,Agrawal:2014aoa, Hamze:2014wca, Lee:2014rba, Kilic:2015vka, Bishara:2015mha,Bhattacharya:2015xha,Calibbi:2015sfa,Baek:2015fma,Chen:2015jkt} have multiple dark matter states by construction, as well as a simple
way to connect the DM states with baryons or leptons that allows the
transfer of asymmetries between the two sectors. Therefore, the SADM
mechanism can be naturally
realized in FDM models. In this work we will use a model of lepton
flavored dark matter to demonstrate how the proposed mechanism works.
We will assume that high-scale leptogenesis~\cite{Fukugita:1986hr}
(see refs.~\cite{Davidson:2008bu,Fong:2013wr} for a review and comprehensive list
of references) generates an asymmetry in the lepton sector, which will
then be transferred to baryons and to the dark sector.

Specifically, consider a model of FDM in which three flavors of SM
singlet Dirac fermions $\left(\chi,\chi^{c}\right)_{i}$ ($i=1,2,3$)
interact with the right-handed leptons of the
SM via a scalar mediator $\phi$, with the interaction given by
\begin{equation}
  \mathcal{L}_{\rm FDM}=\lambda_{i j}  \, \phi\, \chi_i\, e^c_j +{\rm h.c.}
  \label{eq:FDM}
\end{equation}
We will denote the mass of the lightest $\chi$ by $m_{\chi}$ and the
typical mass splitting between the $\chi$ flavors by $\delta m$.

It is worth commenting on the conserved quantum numbers in the presence of the
interaction in equation~\ref{eq:FDM}. Individual lepton numbers $L_{i}$ in the SM can be extended by assigning charges to $\chi_{i}$. We will refer to the extended lepton numbers by $\tilde{L}_i$. Then
$U(1)_{B-\tilde{L}}$ remains unbroken and anomaly-free, except for the
explicit breaking from
heavy right-handed neutrinos. If the coupling matrix
$\lambda_{ij}$ is flavor-diagonal in the charged lepton  and $\chi$
mass basis,
the $U(1)_{\tilde{L}}^{3}$ flavor symmetry is preserved to a good
approximation at low energies, broken only by
the light neutrino mass matrix. The neutrino masses are small
enough to have no effect
on the physics to be discussed here, and will therefore be neglected
from here on. The presence of off-diagonal entries in the couplings
$\lambda_{ij}$ do have interesting phenomenological consequences;
however for the sake of simplicity we will defer the discussion of
these effects to a more detailed study and we will restrict ourselves
to the flavor-universal case with $\lambda_{ij}\equiv
\delta_{ij}\lambda$. Note that there is also a separate $U(1)_{\chi}$
symmetry under which all three $\chi_{i}$ have the same charge and the
mediator $\phi$ has the opposite charge.

As mentioned above, we assume that out-of-equilibrium decays of the
lightest right
handed neutrino $N_1$ generate a net $B-\tilde{L}$ asymmetry in the SM
sector. The comoving quantum numbers
\begin{align}
  \tilde{\Delta}_{i}
  &=\left(B/3-\tilde{L}_{i}\right)/s
  \equiv \Delta_{i}-\Delta Y_{\chi_{i}}
\end{align}
are conserved from the end of leptogenesis down to
scales where neutrino oscillations become important.
Here $s$ is the entropy density, $Y_{\chi_i}=n_{\chi_i}/s$ are the
comoving number densities of dark matter, and
$\Delta_{i}=\left(B/3-L_{i}\right)/s$ are
the conserved comoving quantum numbers in the absence of the dark
sector.
Depending on which linear superposition of the $e$, $\mu$ and $\tau$ flavors
$N_1$ couples to, leptogenesis generates nonzero values for these
conserved quantities, which we will take as the initial conditions for
the SADM mechanism.

\begin{figure}
\includegraphics[width=0.4\textwidth]{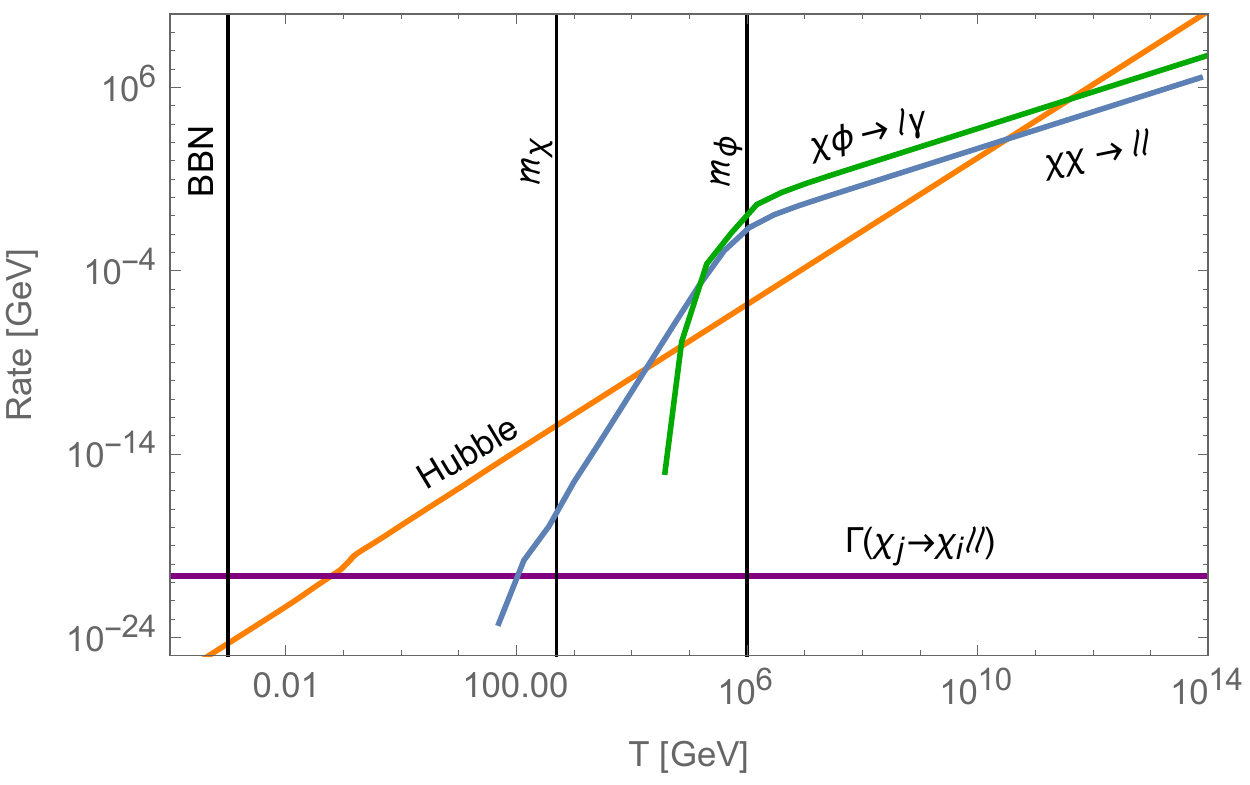}
\caption{Rates of the most important FDM processes and the Hubble scale as a function of temperature for the parameter point defined in the main text.}
\label{fig:timeline}
\end{figure}

Let us now follow the thermal history of the universe from the end of
leptogenesis to lower temperatures. For concreteness we will use a
specific parameter point ($\lambda=0.05, m_\chi=500$~GeV,
$m_\phi=10^6$~GeV, $\delta m=0.4 m_\chi$, $T_{\rm leptogenesis} >
10^{12}~$GeV), and in figure~\ref{fig:timeline} we show for this
parameter point how the rates of the most important processes in the
model compare to the Hubble scale as a function of temperature. With
these values, the FDM interaction of equation~\ref{eq:FDM} goes into
chemical equilibrium after all $N$ have decayed. This is not a
necessary condition for the SADM mechanism to work and merely
simplifies the discussion, as it lets us take initial conditions from
leptogenesis (values of $\Delta_i$, denoted henceforth as
$\Delta^{0}_{i}$) in a modular fashion. If the FDM interaction is
already in equilibrium during leptogenesis one can solve the Boltzmann
equation to track the asymmetries in the two sectors as a function of
time.

As the universe continues to cool down, the asymmetry originally
generated in the left-handed leptons is transferred to the
right-handed leptons (through the SM Yukawas), the baryons (through
sphalerons) and to the $\chi_{i}$ (through the FDM interactions). With
all these interactions in equilibrium, the comoving asymmetries of all
species can be related to the conserved quantities during this epoch
(the $\tilde{\Delta}_{i}$) through equilibrium thermodynamics, with
the constraints that the total hypercharge and the total $U(1)_{\chi}$
number of the universe stay zero. Since individual $\chi$ numbers are
all zero until the FDM interaction goes into equilibrium, the value of
$\left(\tilde{\Delta}_{i}\right)$ just after equilibrium is equal to
the value of $\left(\Delta_{i}\right)-\left(\Delta
Y_{\chi_{i}}\right)$ just before, namely $\Delta^{0}_{i}$.

\begin{figure}
\includegraphics[width=0.4\textwidth]{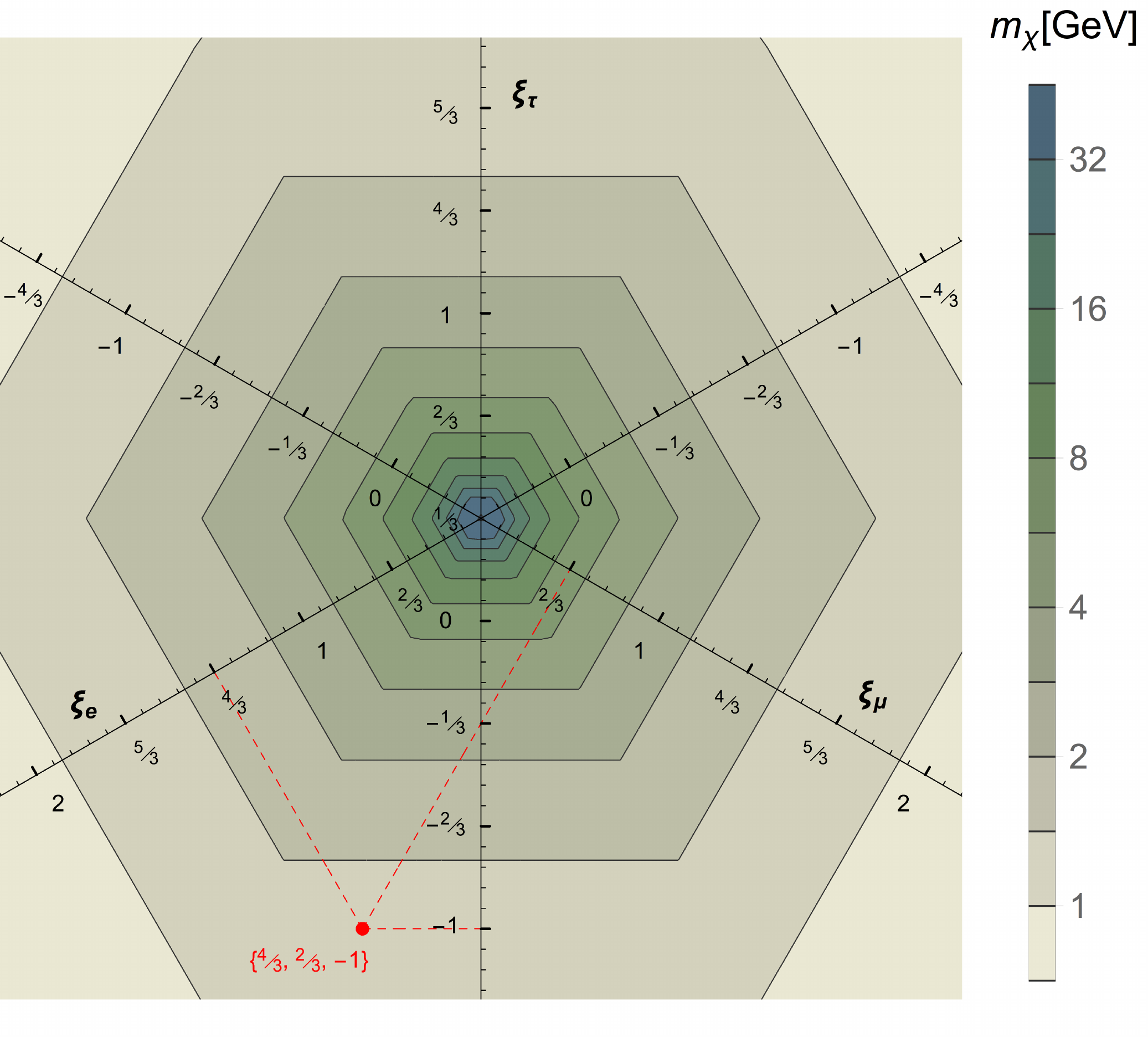}
\caption{The values of $m_{\chi}$ needed to obtain the correct $\rho_{B}$ and $\rho_{DM}$ as the initial lepton asymmetries $\Delta^{0}_{i}$ are
varied subject to the constraint of equation~\ref{eq:constraint}, assuming there is no symmetric component to the relic. The values of $\xi_i \equiv \Delta^{0}_{i} / \Delta Y_{B-L}$ for any point can be read off by drawing perpendiculars to the three axes shown.}
\label{fig:chimass}
\end{figure}

At our parameter point, the next step in the thermal evolution is the
FDM interaction falling out of equilibrium as the temperature drops
below $m_{\phi}$. This decouples the SM and FDM sector asymmetries.
Now the comoving asymmetries $\Delta Y_{\chi_{i}}$ are all separately
conserved, and their values are
given in terms of the initial conditions as
\begin{equation}
\left(\begin{array}{c}\Delta Y_{\chi_{e}}\\\Delta Y_{\chi_{\mu}}\\\Delta Y_{\chi_{\tau}}\end{array}\right)=\frac{2}{15}\left(\begin{array}{ccc}-2 & 1 & 1\\1 & -2 & 1\\1 & 1 & -2\end{array}\right) \left(\begin{array}{c}\Delta^{0}_{e}\\\Delta^{0}_{\mu}\\\Delta^{0}_{\tau}\end{array}\right).
\label{eq:DeltaYs}
\end{equation}
At the same time, the total $B-\tilde{L}$ comoving asymmetry in the SM sector
at early times can be related to the baryon number density $B_{0}$ and entropy density $s_{0}$ today,
\begin{align}
  \Delta Y_{B-\tilde{L}}
  &= \sum_{i} \Delta^{0}_{i}\approx\frac{79}{28} \frac{B_{0}}{s_{0}},
\label{eq:constraint}
\end{align}
which imposes a constraint on the possible initial conditions. From
this point on, the thermal evolution of the SM sector proceeds as
usual.

After the symmetric component of DM annihilates away (through mechanisms
discussed below), the DM relic abundance today is given by
\begin{equation}
\rho_{DM}= m_{\chi}\ s_{0}
\left(|\Delta Y_{\chi_{e}}|+|\Delta Y_{\chi_{\mu}}|+|\Delta Y_{\chi_{\tau}}|\right).
\label{eq:abundance}
\end{equation}
Therefore, the ratio
\begin{equation}
\frac{\rho_{B}}{\rho_{DM}}=\frac{m_{p}}{m_{\chi}}\ \frac{28/79\left(\Delta_{e}^0 + \Delta_{\mu}^0 + \Delta_{\tau}^0\right)}{|\Delta Y_{\chi_{e}}|+|\Delta Y_{\chi_{\mu}}|+|\Delta Y_{\chi_{\tau}}|}
\label{eq:dmmass}
\end{equation}
relates the value of $m_{\chi}$ to observed values of $\rho_{B}$ and $\rho_{DM}$ (with $\rho_{B} / \rho_{DM} = 0.185$~\cite{Ade:2015xua}), given any initial condition $\Delta^{0}_{i}$. This is
illustrated in figure~\ref{fig:chimass}.
Note that $\rho_{B}$ and $\rho_{DM}$
depend on different combinations of the initial conditions.

While for generic initial conditions we expect $m_{\chi}$ to be a few
GeV, both larger and smaller values are possible in the following two
limits: If the leptogenesis mechanism generates almost equal
$\Delta^0_i$ then equation~\ref{eq:DeltaYs} sets the $\Delta
Y_{\chi_{i}}$ to be small, and therefore the DM mass needs to be large
to obtain the right $\rho_{DM}$. On the other hand, if the
leptogenesis mechanism generates large individual asymmetries for the
SM lepton flavors that almost cancel~\cite{MarchRussell:1999ig} (e.g. $\Delta^{0}_{\tau}=
-\Delta^{0}_{\mu}\gg \Delta^{0}_{e}\sim\Delta Y_{B-L}$) then the
denominator in equation~\ref{eq:dmmass} is large, and the DM mass
needs to be small.

\vspace{0.1in}
\noindent

{\it Decays in the dark sector:} If the mass splitting $\delta
m_{ij}\equiv m_{\chi_{i}}-m_{\chi_{j}}$ is less than
$m_{\ell_{i}}+m_{\ell_{j}}$, the decays $\chi_{i}\rightarrow
\chi_{j}+$X can only proceed through $\chi$-flavor mixing or through
strongly suppressed loop processes~\cite{Agrawal:2015tfa}, and the lifetime can be so long
that all three $\chi$ can be treated as stable for practical purposes.
For larger splittings however, the decay
$\chi_{i}\rightarrow\chi_{j}\ell_{i}\bar{\ell}_{j}$ proceeds at tree
level, with
\begin{align}
  \Gamma
  &\simeq
  \frac{\lambda^4 {(\delta m_{ij})}^5}{480 \pi^3 m_\phi^4}.
  \label{eq:decay}
\end{align}

If decays become important before $\chi$-$\bar{\chi}$ annihilations
freeze out, then they depopulate the heavier flavors and the dark
matter abundance is set by the usual symmetric thermal freeze-out.
Therefore, if the relic abundance based on the initial asymmetry is to
survive at late times,
then decays need to happen after annihilations freeze-out, but before
Big-Bang Nucleosynthesis (BBN) in order to avoid early universe
constraints. This is a core requirement of our set up.
It is straightforward to check that this condition is
satisfied at our parameter point. The width of the heavier flavors for
these parameters is
illustrated by the horizontal line in figure~\ref{fig:timeline}.

\vspace{0.1in}
\noindent

{\it Annihilation of the symmetric DM component:}
If
FDM annihilations $\chi_i \bar{\chi}_j \to l_i^{-} l_j^{+}$ are still active below
$T \sim m_\chi$, then they deplete the
asymmetry in the dark sector. Therefore, another core requirement
for SADM is to ensure that the FDM interaction decouples while $\chi$
is relativistic.
This also implies that we need additional interactions which can
annihilate the symmetric component of DM, without
depleting the asymmetry. We
consider the setup, referred from here on as the $Z'$-model, where the $U(1)_{\chi}$ symmetry is gauged with a
coupling $g_{D}$, and where the gauge boson $Z'_{\mu}$ acquires a small
mass $m_{Z'}<m_{\chi}$. The $Z'$ couples to the $\chi_{i}$ in a
flavor-diagonal fashion and leads to efficient
$\chi_{i}$-$\bar{\chi}_{i}$ annihilations, such that the symmetric component of DM annihilates away for $g_{D} \gsim g_{\rm WIMP}$, where
$g_{\rm WIMP}$ is the coupling that leads to the correct relic
abundance for a thermal relic with the same mass.

Since $\phi$ carries a unit charge under $U(1)_{\chi}$ as well as
hypercharge, it leads to kinetic mixing~\cite{Holdom:1985ag, Chun:2010ve} between these groups
\begin{equation}
{\mathcal L}_{\rm mix.}=-\frac{\epsilon}{2} B^{\mu\nu}Z'_{\mu\nu},
\label{eq:mixing}
\end{equation}
where the loop of $\phi$ generates $\epsilon\sim 10^{-3}$ --
$10^{-4}$ for couplings needed to annihilate the symmetric part.
However, other UV contributions to the kinetic mixing can lead to a
larger or smaller value of $\epsilon$. The $Z'$ can
decay to the light SM fermions through the kinetic mixing.

\begin{figure*}[t!]
\includegraphics[width=0.4\textwidth]{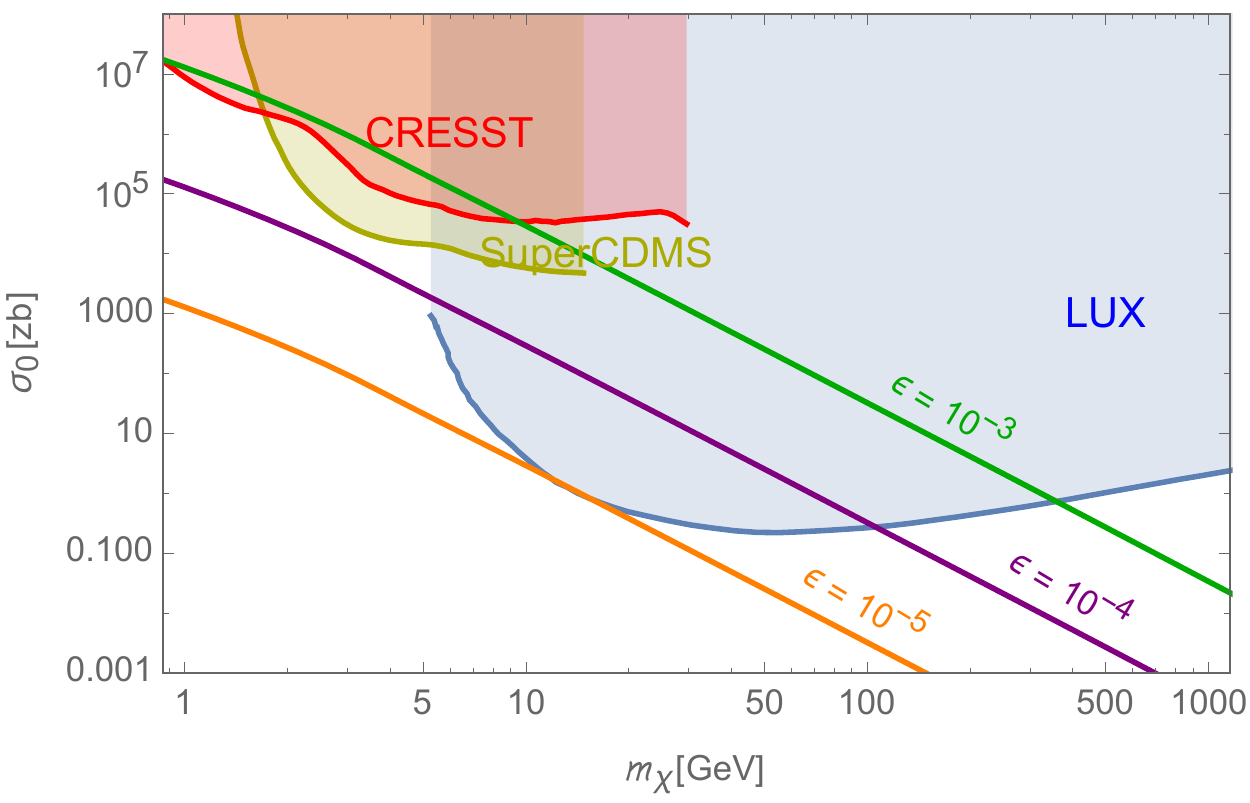}
\qquad
\includegraphics[width=0.4\textwidth]{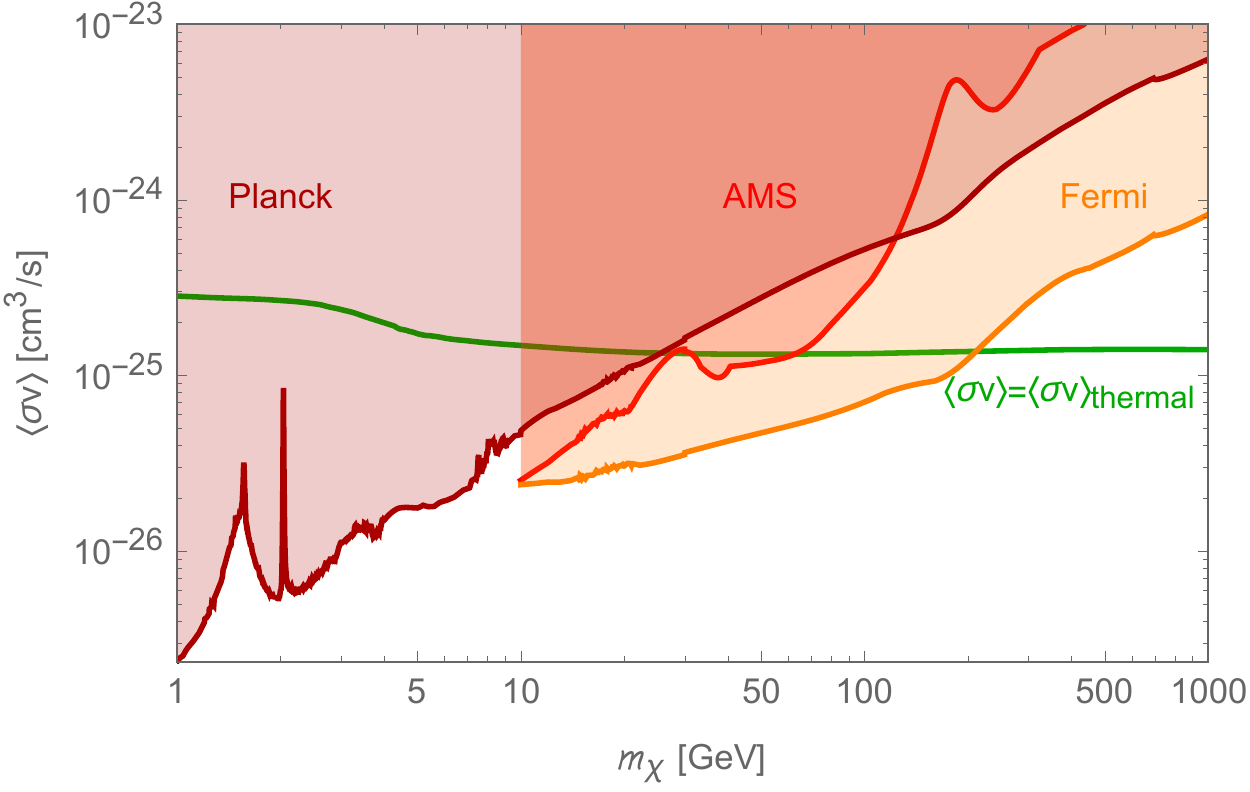}
\caption{
  Constraints on the $Z'$-model.
  \emph {Left:}
  Direct detection constraints
  from LUX~\cite{luxslidesX,Akerib:2015rjg},
  SuperCDMS~\cite{Agnese:2015nto} and CRESST-II~\cite{Angloher:2015ewa}
  for representative values of $\epsilon$ and $g_{D}=g_{\rm WIMP}$.
  \emph{Right:}
  Indirect detection constraints from
  Planck~\cite{Ade:2015xua}, Fermi~\cite{Ackermann:2015zua} and
  AMS~\cite{Accardo:2014lma,Aguilar:2014mma}. For reference we also show the annihilation cross section~\cite{Steigman:2012nb} which gives the correct relic abundance in our model with no asymmetry. $m_{Z'}$ is taken to be $m_\chi/2$ for both plots.
\label{fig:DD}
\label{fig:ID}
}
\end{figure*}

\vspace{0.1in}
\noindent
{\it Experimental Signatures of the $Z'$-model:} If all flavors of $\chi$ are long-lived on cosmological timescales
then there are no annihilations happening today and therefore indirect
detection experiments are not sensitive to this case. If on the other
hand only the lightest flavor survives today, then the DM distribution
is symmetric. Since there is only a lower limit on $g_{D}$, one can
obtain a stronger signal in indirect detection for a given $m_{\chi}$
compared to a WIMP. In particular, the annihilations will take the
form $\bar{\chi}\chi\rightarrow Z'Z'\rightarrow 4f$, where $f$ denotes
SM fermions with $m_{f}<m_{Z'}/2$. Depending on $m_{Z'}$, the leading
constraint from indirect detection may arise from positrons~\cite{Accardo:2014lma,Aguilar:2014mma}, photons~\cite{Ackermann:2015zua}
or CMB measurements of ionization~\cite{Ade:2015xua}. These constraints were considered in
ref.~\cite{ArkaniHamed:2008qn,Slatyer:2015jla, Elor:2015bho}, and they are shown in the right-hand plot of
figure~\ref{fig:ID}.

The $Z'$-hypercharge mixing also gives rise to a signal in direct
detection experiments such as
LUX~\cite{luxslidesX,Akerib:2015rjg},
SuperCDMS~\cite{Agnese:2015nto} and CRESST-II~\cite{Angloher:2015ewa}.
Since tree-level $Z$-exchange is excluded by orders of
magnitude, this translates to a strong constraint on the model
parameters. In the left-hand plot of figure~\ref{fig:DD} we show the bounds in the
$m_{\chi}$-$\sigma_0$ plane for a specific choice of $m_{Z'} =
m_\chi/2$.

Finally, there are also bounds on the model from dark photon searches,
which can be quite stringent for a very light
$Z'$~\cite{Curtin:2014cca,Alekhin:2015byh}.
However for $m_{Z'} \gsim 1$~GeV, the bound for $\epsilon$ is
typically at the $10^{-3}$ level, and generic
values in our model are compatible with this constraint.

We see that direct detection, indirect detection and dark photon
searches provide a complementary set of constraints for the parameter
space of the $Z'$ model.
Light DM with $m_{\chi}\simeq5$ GeV, which can be
obtained from generic initial conditions (see
figure~\ref{fig:chimass}), is unconstrained by direct detection
even for generic values of
$\epsilon$, and can be within reach of future experiments probing light
dark matter. The low $m_{\chi}$ region
is in tension with indirect detection bounds, but
the constraints may be evaded in a modified version of the model, for example if the main
annihilation channel is into neutrinos.
Heavier $m_{\chi}\gsim
O(100~{\rm GeV})$ are unconstrained by either set of bounds.

\vspace{0.1in}
\noindent
{\it Alternative model for annihilating the symmetric part:}
In order to stress the model dependence of some of the bounds considered above, we describe a variation of the model where
DM annihilates via a scalar instead of a $Z'$. In particular, consider
a light real scalar $S$ with the interactions
\begin{align}
  {\mathcal L}_{S}
  &=
  \kappa_{ij} S \chi_{i} \chi^{c}_{j}
  -V(S)
  \, .
\label{eq:S}
\end{align}
Consistent with the $U(1)^{3}_{\tilde{L}}$ global symmetry we will
take $\kappa_{ij}\equiv\delta_{ij}\kappa$. $S$ develops a coupling to
the right-handed SM leptons at one loop through the FDM interaction,
and can therefore efficiently annihilate the symmetric part of the DM
distribution. $S$ does not mix with the Higgs boson until at least the
two-loop order, and even this mixing is suppressed by lepton Yukawa
couplings. Therefore, unlike the $Z'$, tree-level $S$ exchange only
gives a negligible signal in direct detection experiments.
Furthermore, the annihilation channel $\bar{\chi}\chi\rightarrow SS$
is $p$-wave suppressed, which means that even for a fully symmetric
$\chi$ distribution today, indirect detection signals are expected to
be very weak.
Thus, this alternative model is basically unconstrained by the experiments discussed above.

\vspace{0.1in} \noindent {\it Conclusions:} We have studied the
SADM mechanism where for a dark sector with multiple states, the relic
abundance is set by an asymmetry even though the DM number remains zero. If heavier DM states can decay to the
lightest state, then DM is symmetric at late times, whereas otherwise multiple DM components can be present today. This
mechanism is realized naturally in models of FDM. Experimental
signals, if present, arise mainly due to the sector of the model that
is responsible for annihilating the symmetric component of the DM.
We have presented two alternatives for this sector: a
$Z'$-model where $Z'$-hypercharge mixing generically takes place at
the one-loop level, and a scalar model where mixing with the Higgs can
naturally be very small. For the former model there are a number of
experimental constraints from DM searches as well as dark photon
searches, and future experiments should be able to probe a sizable
fraction of the parameter space currently consistent with constraints.
The latter model on the other hand is very difficult to probe
experimentally, and its parameter space is largely unconstrained.

\vspace{0.1in} \noindent {\it Acknowledgments:} We would like to
thank Zackaria Chacko, Roni Harnik, Tim Tait, James Unwin and
Jiang-Hao Yu for useful conversations. The research of  CK, SS and
CT is supported by the National Science Foundation under Grants No.
PHY-1315983 and No.  PHY-1316033.  PA is supported by NSF grant
PHY-1216270.  PA and CK like to thank the Aspen Center for Physics for
its hospitality,
which is supported by National Science Foundation grant PHY-1066293.
PA and CK are also grateful to the Mainz Institute for Theoretical
Physics (MITP) for its hospitality and its partial support during the
completion of this work. SS would like to thank the Perimeter Institute for Theoretical Physics (PI) for partial support during the completion of this work. Research at PI is supported by the Government of Canada through Innovation, Science and Economic Development Canada and by the Province of Ontario through the Ministry of Research, Innovation and Science.

%%%%%%%%%%%%%%%%%%%%%%%%%%%%%%%%%%%%%%%%%%%%%%%%%%%

\bibliography{akst_SADM}

\end{document}